\documentclass[]{aa} 
\usepackage{txfonts}
\usepackage{natbib}
\usepackage{graphicx, xspace}
\usepackage{placeins}

\usepackage[switch]{lineno}
\newcommand{\eg}{{e.g.}}

\newcommand{\ie}{{i.e.}}

\newcommand{\degs} {^\circ}
\newcommand{\num} {$\nu_{\rm max}$\xspace}
\newcommand{\dnu} {$\Delta\nu$\xspace}
\newcommand{\Kepler} {\textit{Kepler}\xspace}
\newcommand{\kepler} {\textit{Kepler}\xspace}

\newcommand{\Lagr}{\mathcal{L}}
\newcommand{\Magr}{\mathcal{M}}
\newcommand{\Dagr}{\mathcal{D}}

\makeatletter

\newcommand{\Rmnum}[1]{\expandafter\@slowromancap\romannumeral #1@}
\makeatother

\usepackage[dvips]{color}
\newcommand{\actionItems}[1]
{{\bf [ToDo: #1]}}

\def\mh{\,$\mu$Hz\xspace}

\def\num{$\nu_{\rm max}$\xspace}
\def\dnu{$\Delta\nu$\xspace}
\def\dn1{$\delta\nu_{01}$\xspace}
\def\dn2{$\delta\nu_{02}$\xspace}

\def\DP{$\Delta\Pi_1$\xspace}
\def\DPobs{$\Delta P$\xspace}

\begin{document}


\title{Testing the asymptotic relation for period spacings from mixed modes
of red giants observed with the \Kepler mission}

\authorrunning{B.~Buysschaert}
\titlerunning{Testing the asymptotic relation for period spacings of red
    giants}

\author{B.~Buysschaert\inst{1,2,3}, 
P.\,G.~Beck\inst{1,4}, 
E.~Corsaro\inst{1,4,5,6}, 
J.~Christensen-Dalsgaard\inst{2}, \\
C.~Aerts\inst{1,7},
T.~Arentoft\inst{2}, 
H.~Kjeldsen\inst{2,8}, 
R.\,A.~Garc{\'{\i}}a\inst{4}, 
V.~Silva~Aguirre\inst{2},
P.~Degroote\inst{1} \\
}
\mail{bram.buysschaert@ster.kuleuven.be} 

\institute{
Instituut voor Sterrenkunde, KU Leuven, Celestijnenlaan 200D, 3001 Leuven, Belgium 
\and Stellar Astrophysics Centre, Department of Physics and Astronomy, Aarhus University, Ny Munkegade 120, 8000 Aarhus C, Denmark 
\and LESIA, Observatoire de Paris, PSL Research University, CNRS, Sorbonne Universit\'es, UPMC Univ. Paris 06, Univ. Paris Diderot, Sorbonne Paris Cit\'e  
\and Laboratoire AIM, CEA/DSM - CNRS - Univ. Paris Diderot - IRFU/SAp, Centre de Saclay, 91191 Gif-sur-Yvette Cedex, France 
\and Instituto de Astrofisica de Canarias, E-38205, La Laguna, Tenerife, Spain 
\and Univ. de La Laguna, Dept. de Astrofisica, E-38206, La\,Laguna, Tenerife, Spain 
\and Dept. of Astrophysics, IMAPP, Radboud University Nijmegen, 6500 GL Nijmegen, the Netherlands 
\and Sydney Institute for Astronomy (SIfA), School of Physics, University of Sydney, 2006, Australia 
}
\abstract
{Dipole mixed pulsation modes of consecutive radial order have been detected for thousands of low-mass red-giant stars with the NASA space telescope \Kepler. Such modes have the potential to reveal information on the physics of the deep stellar interior.}
{Different methods have been proposed to derive an observed value for the gravity-mode period spacing, the most prominent one relying on a relation derived from asymptotic pulsation theory applied to the gravity-mode character of the mixed modes. Our aim is to compare results based on this asymptotic relation with those derived from an empirical approach for three pulsating red-giant stars.}
{We developed a data-driven method to perform frequency extraction and mode identification. Next, we used the identified dipole mixed modes to determine the gravity-mode period spacing by means of an empirical method and by means of the asymptotic relation. In our methodology, we consider the phase offset, $\epsilon_{\mathrm{g}}$, of the asymptotic relation as a free parameter.}
{Using the frequencies of the identified dipole mixed modes for each star in the sample, we derived a value for the gravity-mode period spacing using the two different methods.  These differ by less than 5\%. The average precision we achieved for the period spacing derived from the asymptotic relation is better than 1\%, while that of our data-driven approach is 3\%.}
{Good agreement is found between values for the period spacing derived from the asymptotic relation and from the empirical method. The achieved uncertainties are small, but do not support the ultra-high precision claimed in the literature. The precision from our data-driven method is mostly affected by the differing number of observed dipole mixed modes. For the asymptotic relation, the phase offset $\epsilon_{\mathrm{g}}$ remains ill defined, but enables an more robust analysis of both the asymptotic period spacing and the dimensionless coupling factor. However, its estimation might still offer a valuable observational diagnostic for future theoretical modelling.}


\maketitle

\section{Introduction \label{sec:Introduction}}

Evolved stars that have exhausted their central hydrogen and are now
  performing hydrogen burning in a shell surrounding the helium core are
  generally referred to as red-giant stars or simply red giants \citep[\eg][and
  references therein]{2013osp..book.....C}. In this work, we will concentrate on
  red giants with a mass ranging from $\sim$1$\rm M_{\odot}$ up to $\sim$2$\rm
  M_{\odot}$. These red giants are known to exhibit solar-like oscillations,
which are intrinsically damped and stochastically excited by the convective
motion of the outer layers of the star \citep[][]{1977ApJ...212..243G,
  1986ssds.proc..105D, 1989ApJ...341L.103C, 2010aste.book.....A, tong2015extraterrestrial}.

The seismic analysis of red giants was driven by the photometric
observations of space telescopes, such as \textsc{CoRoT}
\citep{2006cosp...36.3749B, 2009A+A...506..411A} and \Kepler
\citep{2010Sci...327..977B, 2010ApJ...713L..79K}. The vast amount of data of
unparalleled photometric quality have led to numerous substantial breakthroughs
in the asteroseismology of evolved low-mass stars. In particular, a first major
leap forward in our seismic understanding of those stars was achieved by
\citet{2008A+A...478..497K} and \citet{2009Natur.459..398D} with the detection
of non-radial modes in the power-spectral density (PSD) of red giants. This
detection allowed for the application of seismic analyses for
the measurement of the physical parameters describing the oscillations
\citep{2009A+A...506...57D}.

Our understanding of red giants drastically improved after the detection of
their mixed modes \citep{2011Sci...332..205B,2011Natur.471..608B}.  Such
  modes contain information from both the deep, dense interior of the star and its
  convective envelope. 
  Indeed, mixed modes occur due to the coupling between 
  a region in the core where the mode behaves as a gravity
  (\textit{g}-) mode and a region in the envelope with pressure
  (\textit{p}-)mode behaviour; thus a mixed mode probes 
  conditions both in the core and in the envelope.  Exploitation of the detected mixed modes
allowed for the discrimination between two evolutionary stages, the red-giant
branch (RGB; H-shell burning) and the red clump (RC; He-core and H-shell
burning) stars \citep{2011Natur.471..608B, 2011A+A...532A..86M}, as well as the
detection of the rapid core rotation of red giants
\citep{2012Natur.481...55B,2012A+A...548A..10M}.

During its nominal mission, the \Kepler space telescope observed more than
15,000 red giants \citep{2010ApJ...723.1607H, 2011A+A...525A.131H,
  2013ApJ...765L..41S, 2014yCat..22110002H}, before a second reaction wheel
broke down. In addition, three different star clusters containing red giants
were observed, allowing to do ensemble studies \citep{2011A+A...530A.100H,
  2012MNRAS.419.2077M, 2012ApJ...757..190C}. Detailed analyses of the
oscillation spectrum of individual stars are also being carried out
\citep[e.g.][]{2011MNRAS.415.3783D, 2012A+A...538A..73B, 2014A+A...564A..27D, 2015A+A...578A..76C, 2015A+A...579A..83C}. Furthermore, the reliability
of the seismic tools was tested by studying eclipsing binary stars
\citep{2010ApJ...713L.187H, 2013A+A...556A.138F, 2013ApJ...767...82G, 2014A+A...564A..36B,
  2014ApJ...785....5G}.

Meanwhile, various approaches to exploit the mixed modes were presented
  in the literature. The quasi-constant period spacing of mixed modes due to
  their gravity-mode character allows us to characterise the frequency pattern in
  the PSD of given stars. This frequency pattern can be calculated using the asymptotic approximation of high-order low-degree gravity modes for a non-rotating evolved star. The asymptotic period spacing for the periods of such consecutive gravity modes with spherical degree $\ell$ is given as: 
\begin{equation}
\Delta\Pi_{\ell, \rm asym} = \frac{2\pi^2}{\sqrt{\ell(\ell+1)}} \left(\int\limits_{g}^{}\frac{N}{r}\mathrm{d}r\right)^{-1} \ \mathrm{,}
\label{eq:deltaPiBrunt}
\end{equation}
where $N$ is the Brunt-V\"ais\"al\"a frequency and the integration is performed over the \textit{g}-mode propagation cavity $g$ \citep{1980ApJS...43..469T, 2011MNRAS.414.1158C}. Such pure gravity modes have high mode inertias and therefore low photometric amplitudes are expected \citep[\eg][and references therein]{2009A+A...506...57D,2014A+A...572A..11G}.   
 
  As a first estimate for the asymptotic period spacing, $\Delta \Pi_{\ell, \rm asym}$, one can use the observed period
  spacing, $\Delta P$, between mixed modes of the same spherical degree and consecutive radial order 
to deduce the asymptotic period spacing.  \citet{2011Natur.471..608B}, \citet{2011A+A...532A..86M},
  \citet{2012ApJ...757..190C}, and \citet{2013ApJ...765L..41S} used the average of all observed period spacings
   between consecutive dipole mixed modes, $\overline{\Delta P}$, to characterise the
  evolutionary stage of the red giants. However, $\overline{\Delta P}$ is not
  equal to the quasi-constant asymptotic period spacing caused by the gravity-mode
  character of the mixed modes and it therefore does not contain the optimal
  information related to the stellar core.

\citet{2012A+A...540A.143M} proposed a formalism, based on the work by
\citet{1979PASJ...31...87S} \citep[see also][]{1989nos..book.....U}, 
to describe the full observed frequency pattern of
the dipole mixed modes in the oscillation spectrum. This formalism is based on
the asymptotic period spacing, $\Delta \Pi_{\ell, \rm asym}$, 
of pure high radial-order \textit{g}-modes and the coupling between regions of \textit{p}- 
and \textit{g}-mode behaviour to describe the pattern of the dipole mixed modes. The asymptotic relation for the
mixed modes in red giants was recently investigated and verified from detailled stellar and seismic modelling by \citet{2014MNRAS.444.3622J}. In addition, alternative approaches have been recently proposed, e.g. by \citet{2014ApJ...781L..29B}, relying on the mode inertia to determine the period spacing. The challenge for all methods is to determine the value of the period spacing of the dipole modes with the highest reliability possible.

In this work we intend to evaluate the asymptotic relation introduced by \citet{2012A+A...540A.143M} for selected red giants observed by \Kepler. Our work is a first step towards the confrontation of the observationally deduced (asymptotic) period spacing with the value calculated from theoretical stellar models tuned to the star under investigation. Here, we limit to the examination whether the high precision of the derived period spacing reported in the literature is supported by our methodology, which covers a large parameter space.

\section{Observations and sample}
\label{sect:Obs}

Our analysis is focused on red giants observed with the NASA \Kepler space
telescope. The giants investigated in this work were selected from  visual
inspection of thousands of PSDs of the best studied red giants. To be selected,
the stars had to fulfil two strict criteria:
\begin{itemize}
\item a single clear power-excess, showing pulsations with excellent signal-to-noise ratios (SNRs);
\item no visual evidence of rotational splitting in non-radial modes \citep{2003ApJ...589.1009G, 2006MNRAS.369.1281B, 2012Natur.481...55B, 2013A+A...549A..75G}, since that complicates a direct comparison to frequencies computed with theoretical models.
\end{itemize}
Using these selection criteria, we identified the three red giants, KIC\,6928997,
KIC\,6762022, and KIC\,10593078 as good targets for our analysis. The literature
values of the global stellar parameters for these selected stars are presented in
Table\,\ref{tab:KIC_literature}.

The \kepler observations of the three selected red giants were performed in the
long cadence mode with a non-equidistant sampling rate of approximately
29.4\,min, leading to a Nyquist frequency of $\sim$283.5\mh \
\citep{2010ApJ...713L..87J}. The \Kepler dataset covers a time base of
  1470\,d, leading to a formal frequency 
resolution of $0.00787$\mh.
The \Kepler light
curves used in this work were extracted from the pixel data for the individual quarters (Q0-Q17) of
$\sim$90 days each,  following the method described
in \citet{2013...PhD...Bloemen}. The final light curve and the power spectral
density were compiled and calibrated following the procedure by
\citet{2011MNRAS.414L...6G}. Finally, missing data points up to 20\,d were interpolated according to the techniques presented in \citet{2014A+A...568A..10G} and \citet{2015A+A...574A..18P}.

Subsequently, the \Kepler light curves were investigated to determine any
systematics in the PSDs. Since KIC\,6928997 fell on the malfunctioning CCD
module in Q5, Q9, and Q13, no observations were obtained during those
quarters, leading to slightly stronger side lobes in its spectral
window. However, these side lobes were sufficiently weak so that they did not produce significant frequency peaks that would complicate the analysis.

The stars KIC\,6928997 and KIC\,10593078 have previously been studied by \citet{2012A+A...540A.143M, 2014A+A...572L...5M}, who 
reported their asymptotic period spacing, $\Delta \Pi_{\ell, \rm asym}$, and  coupling factor, $q$, based on the asymptotic relation. We introduce the asymptotic relation together with $q$ in Sect.\,\ref{sect:gridsearch}. The values for the asymptotic period spacing are consistent between both studies, but the uncertainties are markedly different: \citet{2012A+A...540A.143M} stated $\Delta\Pi_{1, \rm asym} = 77.21 \pm 0.02$\,s with $q = 0.14 \pm 0.04$, and $\Delta\Pi_{1, \rm asym} = 82.11 \pm 0.03$\,s with $q = 0.13 \pm 0.04$ for KIC\,6928997 and KIC\,105930078, respectively, while \citet{2014A+A...572L...5M} deduced $\Delta\Pi_{1, \rm asym} = 77.2 \pm 1.4$\,s, and $\Delta\Pi_{1, \rm asym} = 82.1 \pm 1.3$\,s. Here, we take a data-driven approach to elaborate on the derivation of those uncertainties.

\begin{table}[t!]
\caption{Fundamental parameters for the stars in our sample.}
\centering
\tabcolsep=8pt

\begin{tabular}{lcccc} 
\hline
\hline
KIC &  $K_{\mathrm{p}}$ (mag)&$\mathrm{T_{eff}}$ (K) & $\mathrm{log \ g}$ ($\mathrm{dex}$) & $\mathrm{[Fe/H]}$\\ 
\hline
6928997		&$11.584 $&$ 4800 \pm 90 $&$ 2.62 $&$ 0.21$\\
6762022		&$11.532 $&$ 4860 \pm 90 $&$ 2.72 $&$ 0.01$\\
10593078		&$11.567 $&$ 4970 \pm 100 $&$ 2.88 $&$ 0.17$\\
\hline
\end{tabular}

\tablefoot{Magnitudes and object identifiers in the \Kepler input catalogue (KIC) are retrieved from the \citet{2009yCat.5133....0K}, other stellar parameters by \citet{2012ApJS..199...30P}.
}
\label{tab:KIC_literature}
\end{table}

\section{Frequency analysis}
\label{sect:FrequencyAnalysis}

The oscillation properties of solar-like pulsators are usually studied from their
PSD diagram. We start our analysis by
determining the global shape of the PSD in Sect.\,\ref{sect:background}. This
allows us to determine the frequency of maximum oscillation power, \num, defined
as the central frequency of the envelope describing the power of the
oscillations. Next, we deduce the large frequency separation, \dnu, in
Sect.\,\ref{sect:dnu}, which describes the equidistant frequency spacing for
pure $p$-modes under the asymptotic description \citep{1980ApJS...43..469T,
  1990ApJ...358..313T}. It is defined as
\begin{equation}
\Delta \nu = \left(2\int \limits_{0}^R \frac{\mathrm{d} r}{c(r)}\right)^{-1} \ \mathrm{,}
\label{eq:dnu}
\end{equation}
where $c(r)$ is the interior sound speed. This parameter is sensitive to the
mean stellar density \citep{1986ApJ...306L..37U}, while \num has been postulated
to scale with the surface gravity $g$ and the effective temperature
$T_{\mathrm{eff}}$ \citep{1991ApJ...368..599B, 1995A+A...293...87K,
  2003PASA...20..203B, 2011A+A...530A.142B}. Therefore, these two quantities
depend on the stellar mass and radius and form the basis of the scaling
relations broadly used in asteroseismology 
to derive fundamental stellar
properties of solar-like pulsators \citep[i.e.][]{2008ApJ...674L..53S,
  2011ApJ...730...63G, 2011ApJ...740L...2S, 2012ApJ...757...99S,
  2013MNRAS.429..423M, 2014ApJ...787..110C}:
\begin{equation}
\nu_{\mathrm{max}} =\left(\frac{M}{M_{\odot}}\right)\left(\frac{R}{R_{\odot}}\right)^{-2} \left(\frac{T_{\mathrm{eff}}}{T_{\mathrm{eff,\odot}}}\right)^{-0.5} \nu_{\mathrm{max,\odot}} \ \mathrm{,}
\label{eq:numax_scaling}
\end{equation}
and
\begin{equation}
\Delta \nu=\left(\frac{M}{M_{\odot}}\right)^{0.5}\left(\frac{R}{R_{\odot}}\right)^{-1.5} \Delta \nu_{\odot} \ \mathrm{.}
\label{eq:deltanu_scaling}
\end{equation}
Solar values are indicated by a subscript $\odot$ and we adopt the
values of \citet{2011ApJ...743..143H} for
$\nu_{\mathrm{max,\odot}}$ and $\Delta \nu_{\odot}$, which are $3150\,\mu$Hz and
$134.9\,\mu$Hz, respectively.

In a subsequent step, we extract and identify the individual oscillation
modes. To characterise the parameters of each mode in the PSD, we have
constructed a semi-automated pipeline. The details on the methods adopted for
the extraction and identification of the different modes are presented in
Sect.\,\ref{sect:pipeline_peakbagging}. The results obtained for the selected
sample of red giants are given in
Sect.\,\ref{sect:pipeline_peakbagging_results}, and the detailed analysis of the
observed dipole mixed modes is further discussed in
Sect.\,\ref{sect:Period_spacing}.

\subsection{Determination of the background \label{sect:background} }

The overall shape of the power spectrum of a solar-like oscillator is generally
described by a combination of power laws to describe the granulation background signal
and a Gaussian envelope to account for the
position of the oscillation power excess
\citep[\eg][]{1985ESASP.235..199H,2010A+A...509A..73C, 2010A+A...522A...1K}.
The global model of the PSD, expressed as a function of the frequency $\nu$, is
given by
\begin{equation}
\Magr_{\mathrm{PSD}}(\nu) =  \left[P_{\mathrm{gran}}(\nu) + P_{\mathrm{Gauss}}(\nu)
\right] \cdot R(\nu) + W\ ,
\label{eq:pipeline_background}
\end{equation}
where each term is defined below. 

The term corresponding to the granulation signal, $P_\mathrm{gran}(\nu)$, is
expressed as a sum of $s$ different Lorentzian-like profiles
\begin{equation}
P_{\mathrm{gran}}(\nu) = \sum\limits_{i=1}^{s} \frac{2 \pi a_i^2 / b_i}{1 + \left(\nu / b_i \right)^{c_i}}  \ \mathrm{,}
\label{eq:pipeline_granulation}
\end{equation}
where each power law is characterised by its amplitude, $a_i$, its
characteristic frequency, $b_i$, and its slope, $c_i$. In general two or three
different terms are used to describe the granulation contribution, since the
granulation activity occurs on different timescales. 

\citet{1985ESASP.235..199H}  introducted a Lorentzian profile to describe the 
granulation signal of the Sun. More recently, \citet{2010A+A...509A..73C} and
\citet{2010A+A...522A...1K} found a super-Lorentzian profile (\ie\ a
Lorentzian with a slope larger than two) to be more appropriate. 
\citet{2014A+A...570A..41K} provided a detailed overview of the determination of
the shape of red-giant PSDs and suggested that the description with a slope
set to four is favourable. However, in our analysis we choose to keep
the slope as a free parameter within the range two to four, to allow more
degrees of freedom during the fitting phase for a better fit quality.

The shape of the power excess in the PSD is traditionally described with a Gaussian function
\begin{equation}
P_{\mathrm{Gauss}}(\nu) = 
P_g \exp{\left(-\frac{\left(\nu - \nu_{\mathrm{max}}\right)^2}{2\sigma_{g}^2}\right)},
\label{eq:gauss}
\end{equation}
while the instrumental noise $W$ in Eq.\,(\ref{eq:pipeline_background}) is assumed to
be constant.
Aside from the abovementioned contributions to the global model of the PSD, the
sampling effects of the dataset need to be considered, since the discretisation
of the signal may reduce the power of both the oscillation and the granulation
contributions \citep{2013ApJ...767...34K, 2014A+A...570A..41K}. This effect is
taken into account by including the response function, $R(\nu)$,
\begin{equation}
R(\nu) = \mathrm{sinc}^2\left(\frac{\pi \nu}{2 \nu_{\mathrm{Nyq}}}\right)  \ \mathrm{,}
\label{eq:pipeline_response}
\end{equation}
where $\nu_{\mathrm{Nyq}}$ is the Nyquist frequency of the \Kepler long cadence data. 

Estimates of the different parameters describing the background given by
Eq.\,(\ref{eq:pipeline_background}) are first obtained by means of a least-squares
(LS) minimisation technique. We subsequently refine these estimates by using a
Bayesian MCMC (Markov Chain Monte Carlo) algorithm
\citep[\eg][]{2004cond.mat.10490B}. This technique minimises the following
log-likelihood function, $\Lagr(\Theta)$ \citep{1986ssds.proc..105D,
  1990ApJ...364..699A},
\begin{equation}
\Lagr(\Theta) = \sum \limits^{}_{k} \left\{\log \left(\Magr(\Theta; \nu_k)\right) + \frac{\Dagr(\nu_k)}{\Magr(\Theta; \nu_k)}\right\}  \ \mathrm{,}
\label{eq:bayesian_likelihood}
\end{equation}
where $\Dagr(\nu_k)$ is the data for a certain frequency region $k$ and
$\Magr(\Theta;\nu_k)$ is the model with the vector parameter $\Theta~=~\left(\Theta_1, \Theta_2, \dots, \Theta_n \right)$, having $n$ dimensions.  The
logarithm ensures a high numerical stability. In the current case, the data we
wish to describe, $\Dagr$, corresponds to the observed PSD and $\Magr$ the general shape
of the PSD. We use uniform priors for all variables $\Theta_i$ defining the
model $\Magr(\Theta;\nu_k)$, hence we set an upper and lower boundary to each
parameter with a uniform probability distribution.

\begin{figure}[t!]
\centering
\includegraphics[width=\columnwidth]{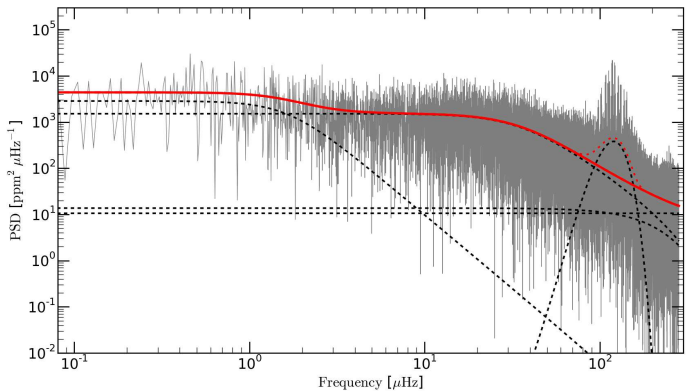}%
\caption{Power density spectrum of the \Kepler photometry of
    KIC\,6928997. We represent the individual components describing the
    background of the PDS by the black dotted lines, their joint effect by the
    red full line, and the power excess by the red dashed line. Only the solid
  red line is used as background during the extraction of the oscillations.}%
\label{fig:KIC6928997_background}%
\end{figure}

Upon convergence of the Bayesian MCMC routine, the marginal posterior
probability distribution is determined for each fitting parameter. We accept the
median value of this distribution as the true value for a given parameter 
in order to capture skewed distributions. Uncertainties on each
$\Theta_i$ are furthermore extracted from the probability distribution. We
present the determined description of the shape of the PSD of KIC\,6928997 in
Fig.\,\ref{fig:KIC6928997_background} and also indicate the aforementioned
individual components.

\subsection{Determination of the large frequency separation}
\label{sect:dnu}

In a second step, we deduce the global large frequency separation \dnu. To do
so, we calculate the autocorrelation function (ACF) of the PSD over a predefined
frequency region centred around \num. The region for the ACF method is defined
by using a multiple of the estimate for \dnu, which is derived from a scaling
relation with the previously derived value of \num. We use the description by
\citet{2010ApJ...723.1607H}, because it has been calibrated on a large sample of
stars. It is given by
\begin{equation}
\Delta \nu_{\mathrm{estimate}} = \left(0.263 \pm 0.009\right) \ \nu_{\mathrm{max}}^{0.772\pm0.005} \ \mathrm{.}
\label{eq:dnu_estimate}
\end{equation}
Here, $\Delta \nu_{\mathrm{estimate}}$ represents the estimated value for \dnu
through the scaling relation. The frequency region $\nu_{\mathrm{max}} \pm
2\Delta \nu_{\mathrm{estimate}}$ is then passed to the ACF routine to determine
a more precise value. We choose to include only this region since we are less
sensitive to the variations of $\Delta\nu$ with varying $\nu$ in that way.

Next, we determine the maximum of the ACF in a region around
$\Delta\nu_{\mathrm{estimate}}$ and further refine it by fitting a Lorentzian
profile. Through this approach, the obtained value for \dnu is not sensitive to
the frequency resolution of the ACF.  Similar to the PSD model fitting process,
the initial estimates on the parameters of the Lorentzian profile are calculated
by means of  LS minimisation. These are finally passed to the
Bayesian MCMC algorithm, again assuming a uniform prior on the fitting parameters and
the likelihood function defined in Eq.\,(\ref{eq:bayesian_likelihood}).

\subsection{Extraction of the oscillation modes}
\label{sect:pipeline_peakbagging}

Once \num and \dnu are accurately determined, we extract and identify the
individual oscillation modes. We choose to fit the modes one \textit{radial-mode
  order} at a time, \ie\ we only consider all significant modes between two
consecutive radial-mode orders $n_{\mathrm{p}}$ and $n_{\mathrm{p}} + 1$, instead of
performing a global fit. We define this radial-mode order, $n_{\mathrm{p}}$, as
the one of the radial modes and will use it during the mode identification
process. It differs from the radial order of the mixed dipole modes, which we
call {\it mixed-mode order\/} and denote as $n_{\mathrm{m}}$. The mixed-mode
order is dependent upon $n_{\mathrm{p}}$ and upon the radial order of the pure
gravity modes, $n_{\mathrm{g}}$ \citep[see
\eg][]{2012A+A...540A.143M}. Performing the fitting in a small frequency range
leads to fast convergence, since the individual modes are well separated and
there are less free parameters compared to fitting the full PSD at once.

A signal-to-noise criterion determines the significance of a given oscillation mode. We
calculate this SNR by dividing the PSD by its global shape,
Eq.\,(\ref{eq:pipeline_background}), while excluding the Gaussian term,
Eq.\,(\ref{eq:gauss}). This general shape corresponds to the solid red line in
Fig.\,\ref{fig:KIC6928997_background}, while the power-excess is represented by
the dashed red line. Mode peaks are considered significant when their SNR is
above 7 times the average SNR in the frequency range of that particular radial-mode
order.

The profile of solar-like oscillation modes in a PSD is represented by a
Lorentzian \citep{1988ApJ...328..879K, 1990ApJ...364..699A}, described by
\begin{equation}
P_{\mathrm{mode}}(\nu) = \frac{A^2 / \pi \Gamma}{1 + 4 \left(\frac{\nu - \nu_0}{\Gamma}\right)^2} \ \mathrm{,}
\label{eq:Lorentzian_peak_bagging}
\end{equation}
where the amplitude, the full width at half maximum (FWHM) and the central
frequency of the Lorentzian profile are given by $A$, $\Gamma$, and $\nu_{0}$,
respectively. The profile of $k$ significant oscillation modes
in a given radial-mode order is then given by
\begin{equation}
\Magr_{\mathrm{order}}(\nu) = W + R(\nu) \cdot \left[\ P_{\mathrm{gran}}(\nu) +  \sum \limits^{k}_{j=1} P_{\mathrm{mode}, j}(\nu) \right] \ \mathrm{.}
\label{eq:Lorentzian_order}
\end{equation}
Both the white noise contribution, $W$, and the granulation contribution,
$P_{\mathrm{gran}}(\nu)$, have been determined during the fit to the
PSD. Therefore, we keep them fixed during the fitting process per radial-mode order,
i.e. only the parameters influencing the individual significant frequencies are
varied. Again, we consider the response function, $R(\nu)$, to account for the
discrete sampling of the photometric signal.

The developed semi-automated \textit{peak-bagging\/} algorithm consists in total of three different steps. First, all significant oscillation modes in a given radial-mode order are fitted with Lorentzian profiles superimposed on the derived background model of the PSD in an automated manner. Second, an interactive fitting step allows the user to have more influence on the fitting process of the individual modes. This is sometimes necessary when the LS minimisation does not converge properly for the very long-lived $g$-dominated mixed modes. Finally, once the initial guesses for the parameters of the fit are sufficient, a Bayesian MCMC algorithm determines the marginal posterior probability distribution for each fitting parameter of an individual mode. The likelihood for the Bayesian fit per radial-mode order is again described by Eq.\,(\ref{eq:bayesian_likelihood}). We adopt uniform priors for both the central frequency and the amplitude of a given peak, while a modified Jeffreys' prior \citep{2011A+A...527A..56H} is used for the FWHM of the Lorentzian profiles. 

The mode identification was performed for the retrieved peaks by using a dimensionless reduced phase shift $\theta$ defined from a frequency \'echelle diagram as 
\begin{equation}
\theta = \left( \nu / \Delta \nu \right) - \left(n_{\mathrm{p}} + \epsilon \right)
\label{eq:theta}
\end{equation}
and having a value in the interval {$-0.2 \leq \theta < 0.8$}. Here, $\nu$ and $n_{\mathrm{p}}$ are the frequency and the radial-mode order, respectively, and $\epsilon$ is a small constant that occurs from an asymptotic approximation for the mode frequencies. This constant can be approximated using the scaling relation presented in \citet{2011A+A...525L...9M}, and updated by \citet{2012ApJ...757..190C}, which leads to radial modes having $\theta \approx 0.00$ and quadrupole modes having $\theta \approx -0.12$ \citep[see also][]{1980ApJS...43..469T, 2012A+A...548A..10M}. Dipole \textit{p}-modes, on the other hand, have $0.2 \leq \theta < 0.8$ and $\ell=3$ modes have $0.1 < \theta < 0.2$. We choose to estimate $\epsilon$ for each radial-mode order from the radial modes assuming their $\theta$ to be exactly zero and approximating $n_{\mathrm{p}}$ as $\lfloor \nu_{\ell=0} / \Delta \nu \rfloor -1$, instead of using the empirical scaling relation.

\subsection{Results}
\label{sect:pipeline_peakbagging_results}

The derived values for the global asteroseismic properties \num and \dnu
obtained by using the Bayesian MCMC methods are presented in
Table\,\ref{tab:results_frequency}, including their $1\sigma$
uncertainties corresponding with an 68\% probability that the true value
is included in the overall interval.

\begin{table}[t!]
\caption{{Determined} seismic parameters of the stars in our sample.}
\centering
\begin{tabular}{lcc} 
\hline
\hline
		KIC					&$\nu_{\mathrm{max}}$ ($\mathrm{\mu Hz}$)&$\Delta \nu$ ($\mathrm{\mu Hz}$)\\
\hline
6928997			&$	119.13	^{+0.39}	_{-0.35}	$&$	10.015_{-0.005}^{+0.005}			$\rule{0pt}{3ex}\\
6762022			&$	41.02	^{+0.18}	_{-0.22}	$&$	4.455_{-0.009}^{+0.010}			$\rule{0pt}{3ex}\\
10593078			&$	206.98	^{+0.25}	_{-0.34}	$&$	15.428	^{+0.019}	_{-0.020}	$\rule{0pt}{3ex}\\
\hline
\end{tabular}
\tablefoot{{The frequency of the oscillation power excess \num, and the large
    frequency separation \dnu, obtained with the Bayesian MCMC technique. Uncertainties noted here are $68\,\%$ confidence intervals.}}
\label{tab:results_frequency}
\end{table}

We find significant oscillation modes in regions in-between six consecutive
radial modes, corresponding to five different radial-mode orders $n_{\mathrm{p}}$.
Therefore, we consistently restrict the analysis at most to the central five
radial-mode orders throughout this work. Additional care is
taken to only obtain frequencies that could unambiguously be identified, \ie\ no
dipole mode was considered outside the expected region. Although other work
often takes such modes into account, increasing the total number of frequencies
available, we choose to reject them. For KIC\,6762022, 
significant dipole modes only occurred 
in its three central radial-mode regions so we restricted to this regime for
this star.

\section{Period spacing analysis}
\label{sect:Period_spacing}

\begin{figure}[t!]
\centering
\includegraphics[width=\columnwidth, height = 0.33\textheight]{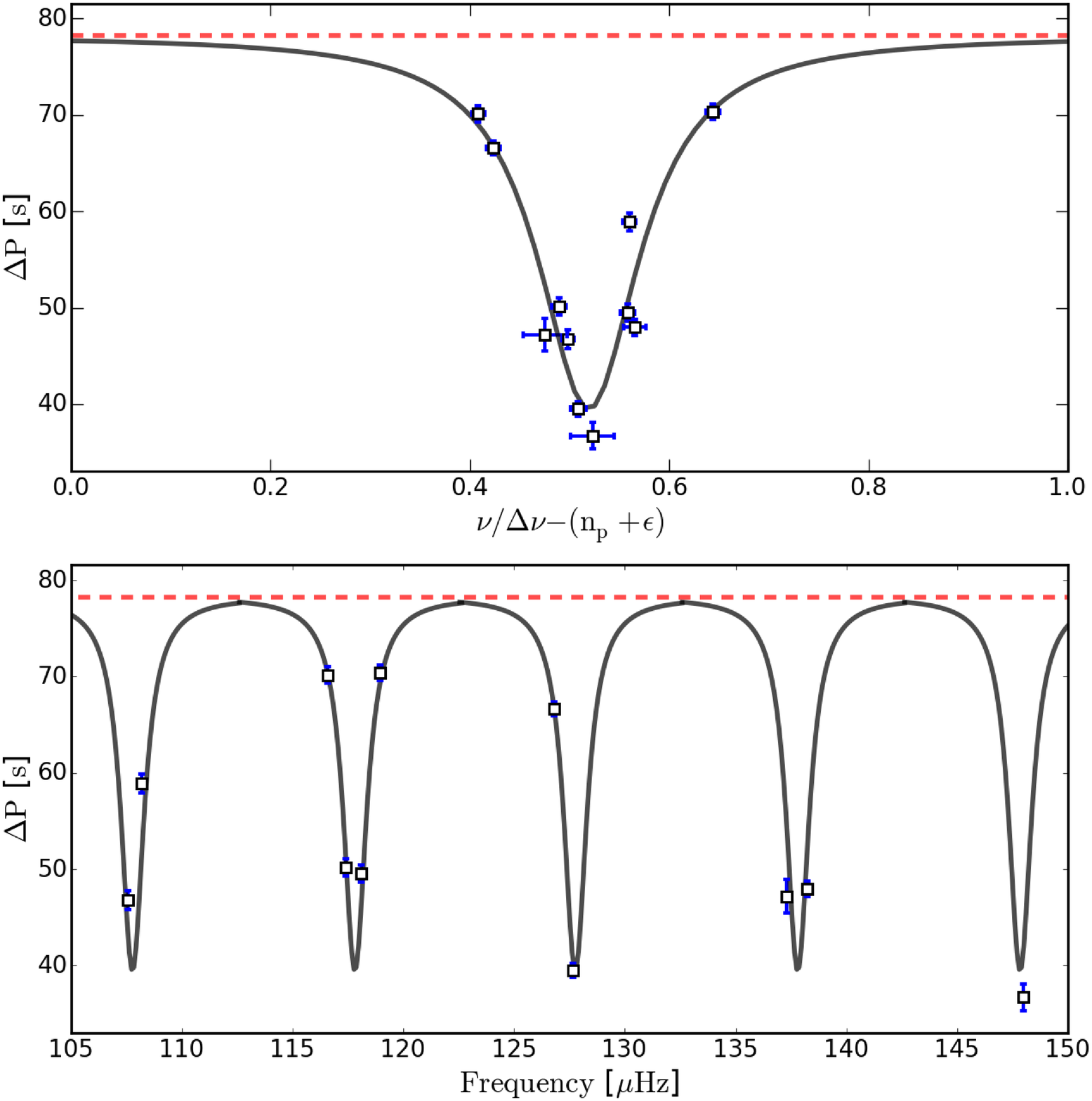}%
\caption{Derivation of $\Delta \Pi_{1, \rm emp}$ for KIC\,6928997 using the empirical approach of fitting \DPobs, yielding $\Delta \Pi_{1, \rm emp} = 81.14 \ \mathrm{s}$ (red dashed). The fit is done for the phase shift $\theta$ (\textit{top}) and subsequently expanded to $\nu$ (\textit{bottom}), using the inverse relation of Eq.\,(\ref{eq:theta}).
  \label{fig:KIC6928997_deltaP}}
\end{figure}

The dipole mixed modes in evolved stars take on an
  acoustic-mode nature in the outer envelope and a gravity-mode nature in the
  deep interior \citep{2001MNRAS.328..601D, 2004SoPh..220..137C,
    2009A+A...506...57D, 2010ApJ...721L.182M}. This causes their period spacing
  to deviate from the constant value expected for pure $g$-modes in the
  asymptotic approximation. The dipole mixed modes turn out to be detectable at
  the stellar surface by means of high precision photometry
  \citep{2011Sci...332..205B, 2011Natur.471..608B,2011A+A...532A..86M}.

The period difference (expressed in seconds) between two consecutive dipole
  mixed modes is formally named the observed period spacing
  and is defined as
\begin{equation}
\Delta P = \frac{1}{\nu_{n_{\mathrm{m}}}} - \frac{1}{\nu_{n_{\mathrm{m}}+1}}\,.
\label{eq:deltaPobs}
\end{equation}
The average of all observed period spacings, $\overline{\Delta P}$ is a
good indicator of the evolutionary stage of a given star
\citep[e.g.][]{2011Natur.471..608B,2011A+A...532A..86M,2012ApJ...757..190C} and
is reported in Table\,\ref{tab:results_DP}. 

The observed period spacing measured from consecutive mixed modes, $\Delta P$,
can be used to infer the value of the asymptotic period spacing of dipole mixed modes
\DP, which is given analytically by Eq.\,(\ref{eq:deltaPiBrunt}) for any
spherical degree $\ell > 0$. Here, we test two different methods to derive \DP
and explore their reliability, without considering frequency-dependent
variations of \DP caused by structural glitches in the core of the star
\citep{2015ApJ...805..127C}.

\subsection{Lorentzian fitting to $\Delta P$}
\label{sect:LorentzModulation}

To obtain a first estimate for the value of the asymptotic period spacing \DP for a
given red-giant star, we use an empirical approach
\citep[see][]{2012ASPC..462..200S}. 
This approach captures the mixed character
due to the pressure and the gravity nature of the dipole modes seen in the
observed period spacings, $\Delta P$, related to the mode bumping
\citep{2010Ap+SS.328..259D}. When mixed modes show a very strong gravity
character, $\Delta P$ remains fairly constant and close to the value of
\DP. However, when the character of the mixed modes becomes more pressure-like,
a lower $\Delta P$ is expected and observed. Thus, the behaviour of the mixed
modes can be captured by a convolution of a flat continuum, accounting for \DP
in case of pure $g$-modes, and a Lorentzian profile with negative height,
taking the mixed mode nature into account by reducing the strength of the
$g$-mode character of the mixed modes.

We choose again to work with the previously defined dimensionless reduced phase shift $\theta$, since
it allows for a more stable fitting procedure. An average phase shift
\begin{equation}
\theta_{\rm spacing} = \frac{\theta_{n_{\mathrm{m}}} + \theta_{n_{\mathrm{m}}+1}}{2} \ \mathrm{,}
\label{eq:theta_average}
\end{equation}
is assigned for each observed period spacing $\Delta P$. This enables us to describe the profile for the mixed mode period spacings as:
\begin{equation}
\Magr_{\rm emp}(\theta_{\rm spacing}) = \Delta\Pi_{\rm emp} - \frac{H}{1 + 4 \left(\frac{\theta_{\rm spacing} - \theta_0}{\Gamma}\right)^2} \ \mathrm{,}
\label{eq:DP_empirical}
\end{equation}
where $H$ is the height of the Lorentzian profile centred at $\theta_0$ with a width $\Gamma$. A LS minimisation technique is adopted to perform the process of fitting $\Magr_{\rm emp}$ to the observed period spacings in terms of the parameters $\Delta \Pi_{\rm emp}$, $H$, $\Gamma$, and $\theta_0$. Estimates on the uncertainties on the individual free parameters are deduced by means of a Monte Carlo approach. We randomly perturb the extracted the extracted dipole frequencies 25,000 times within their respective uncertainties and determine their $\Delta P$ and $\theta_{\rm spacing}$. For the perturbation, we assume normally distributed errors. The fitting process is repeated on each iteration and the scatter on the final set of fitting parameters is then an indication on their uncertainties. The results for this method are reported for the three stars in Table\,\ref{tab:results_DP} as the empirical values $\Delta \Pi_{1, \rm emp}$.

\subsection{Exploration of the asymptotic relation}
\label{sect:gridsearch}

\citet{2012A+A...540A.143M} proposed to derive \DP by solving the equations of
\citet{1979PASJ...31...87S} and formulated the approximation for a frequency of
a dipole mode $\nu_{\rm m}$ as
\begin{equation}
\nu_{\rm m} = \nu_{\mathrm{n_p},\ell=1} + \frac{\Delta \nu}{\pi} \mathrm{arctan}\left[ q \ \mathrm{tan}\, \pi \left(\frac{1}{\Delta \Pi_{1, \rm asym} \nu_{\rm m}} - \epsilon_{\mathrm{g}} \right)\right] \ \mathrm{.}
\label{eq:asymptotic_relation}
\end{equation}
The frequencies of the gravity modes are coupled to the frequency of the pure
$p$-mode $\nu_{\mathrm{n_p}}$. The dimensionless coupling factor $q$
describes the strength of the coupling between the gravity mode cavity and the
pressure mode region. It typically has a value between $0.1$ and $0.3$
\citep{2012A+A...540A.143M}. The constant $\epsilon_{\mathrm{g}}$ is a phase
offset that ensures a proper behaviour for the \textit{g}-mode periods in the case of
weak coupling. \citet{2012A+A...540A.143M} assumed this constant to be zero.
Here, we explore the possibility of varying this phase offset rather than keeping it constant.

Because the asymptotic relation is an implicit
equation for $\nu_{\rm m}$, we solve Eq.\,(\ref{eq:asymptotic_relation}) by
means of a geometrical technique, oversampling the observed frequency 
resolution a 1000 times, as described by
\citet{2013...PhD...Beck}. Second-order asymptotics are considered for the pure pressure modes. Different
approaches have been proposed to derive $\Delta \Pi_{1, \rm asym}$ with the
asymptotic relation from observations. Typically, a LS minimisation with an
initial guess close to the expected value of $\Delta \Pi_{1, \rm asym}$ is used
to search in a narrow range of solutions \citep[\eg][]{2012A+A...540A.143M,
  2014A+A...572L...5M}.

Our aim is to investigate in depth the solution of \DP, using the asymptotic relation, by exploring a
tri-dimensional parameter space for ($\Delta \Pi_{1, \rm asym}$,\,$q$,\,$\epsilon_{\mathrm{g}}$),
thus allowing us to understand the reliability of the final estimate
of the asymptotic dipole period spacing. This is accomplished by means of a
grid-search method, where we vary the parameters \DP and $q$ over a wide range
of values, while $\epsilon_{\mathrm{g}}$ is varied between 0 and 1. As far as we are aware, it is the first time that the phase offset, $\epsilon_{\mathrm{g}}$, of the asymptotic relation is considered a free parameter.

The fit quality for a combination of values of ($\Delta \Pi_{1, \rm asym}$,\,$q$,\,$\epsilon_{\mathrm{g}}$) is
quantified by means of a $\chi^2$ test, where we compute the difference between
the predicted asymptotic frequencies and those observed. The adopted $\chi^2$ is
defined as
\begin{equation}
\chi^2_{\mathrm{grid}} (\Delta \Pi_1,\,q,\,\epsilon_{\mathrm{g}}) = 
\frac{1}{N-4}\sum \limits^{N}_{i} \left(\frac{\nu_{\ell=1,i,\mathrm{obs}} - \nu_{\ell=1,i,\mathrm{asym}}}{\sigma(\nu_{\ell=1,i,\mathrm{obs}})}\right)^2 \ \mathrm{,}
\label{eq:asymptotic_relation_chi}
\end{equation}
where $\nu_{\ell=1,i,\mathrm{obs}}$ is the frequency of the $i$-th observed
dipole mixed mode and $\sigma(\nu_{\ell=1,i,\mathrm{obs}})$ its corresponding
standard deviation estimated by the Bayesian MCMC fit. The related frequency from the
asymptotic relation, the closest one to the $i$-th observed mixed mode
$\nu_{\ell=1,i,\mathrm{asympt}}$, is calculated following 
\citet{2013...PhD...Beck} and $\chi^2_{\mathrm{grid}}$ is normalised by $N-4$
degrees of freedom, where $N$ is the number of observed dipole mixed modes for a
given red-giant star.

The value of $\Delta \Pi_{1, \rm emp}$, determined in
  Sect.\,\ref{sect:LorentzModulation}, acts as a starting point for the $\Delta
  \Pi_{1, \rm asym}$ axis in the tri-dimensional parameter space analysis. We
  let this parameter vary up to $\pm 10\%$ of the initial guess $\Delta \Pi_{1,
    \rm emp}$, and sample with a resolution of 0.02\,s (0.04\,s for RC
  stars). Solutions with a coupling factor $q$ ranging from 0.01 up to 0.51,
  with a resolution of 0.005, were considered. The phase offset
  $\epsilon_{\mathrm{g}}$ spanned from 0 to 1, with a step of 0.0025, noting
  that any values smaller than zero or larger than one will behave similarly as
  those in the used range due to the periodicity of the tangent function. As
  such, we construct a grid of several millions of points, with dimension  
(500,100,400), each grid point representing a unique combination of ($\Delta \Pi_{1, \rm
    asym}$,\,$q$,\,$\epsilon_{\mathrm{g}}$). At each meshpoint, we compare the
calculated frequencies using the asymptotic relation with those extracted from
the PSD to obtain the most probable solution. This best set of values for
  $\Delta \Pi_{1, \rm asym}$, $q$, and $\epsilon_{\mathrm{g}}$ is reported in
  Table\,\ref{tab:results_DP}.

To study each dimension in our tri-dimensional parameter space in more
  detail, we use marginal distributions.  In practice, we consider a minimum $\chi^2_{\mathrm{grid}}$ per gridpoint along
  one dimension to reduce the dimensionality and by marginalising over this
  dimension we assess the correlation between the two remaining
  dimensions. Applying this technique over two dimensions enables us to
  construct confidence intervals for the third, remaining dimension, since it
  represents the distribution of that dimension, similar to the computation of
  the marginal posterior probability in a Bayesian MCMC analysis.

We construct the confidence intervals by relying on the theory of $\chi^2$ statistics using $\chi^2$ tests, i.e. the $\chi^2_{\alpha}$ value for an $\alpha\,\%$ confidence interval. This $\chi^2_{\alpha}$ level corresponds to a $\chi^2_{\mathrm{grid}}$ value indicating the found $\chi^2_{\mathrm{grid}}$ minimum is the real minimum with an $\alpha\,\%$ certainty and is computed as
\begin{equation}
\chi^2_{\alpha} = \frac{\chi^2_{\alpha,k}\ \cdot \chi^2_{\mathrm{grid,\ min}}}{k} \ \mathrm{,}
\label{eq:chi_alpha}
\end{equation}
with $k = N-4$ degrees of freedom,  $\chi^2_{\mathrm{grid,\ min}}$ the value for
  the optimal solution, and $\chi^2_{\alpha,k}$ the tabulated value for an
  $\alpha\,\%$ inclusion of the cumulative distribution function of a $\chi^2$
  distribution with $k$ degrees of freedom. We divide by the degrees of freedom,
  since the $\chi^2$ test in Eq.\,(\ref{eq:asymptotic_relation_chi}) contains a
  sum of multiple distributions. The multiplication by $\chi^2_{\mathrm{min}}$
  accounts for the difference between the optimal value and the expected value
  of 1. The parameter ranges in the 1D marginal distributions are then taken as the
  confidence interval for that parameter.


\subsection{Results}
\label{sect:DP_results}

\begin{figure*}[t!]
\centering
\includegraphics[width=\textwidth]{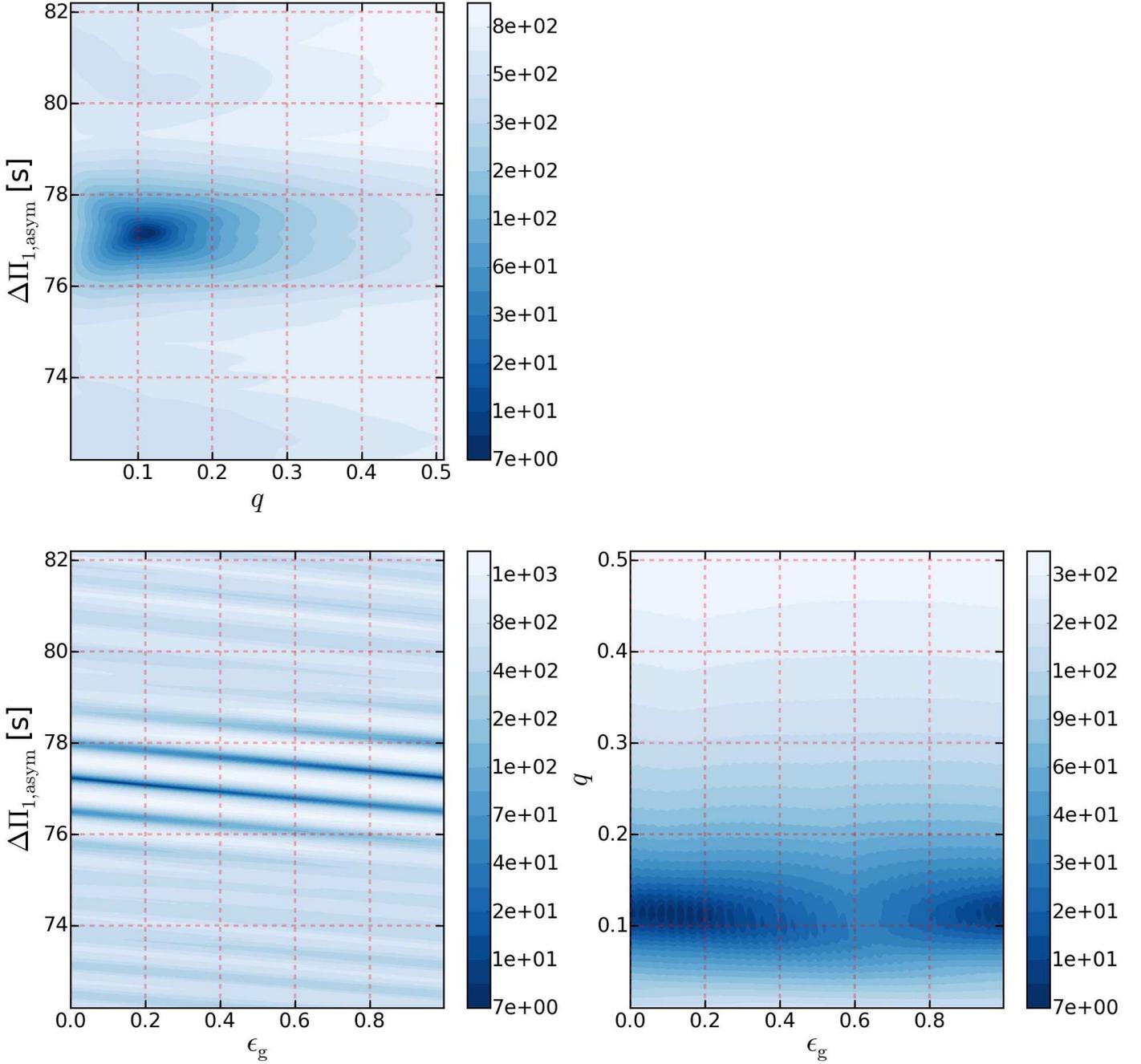}%
\caption{Correlation maps indicating the $\chi^2_{\mathrm{grid}}$ as a
    colour map for KIC\,6928997, as deduced from the asymptotic relation
    Eq.\,(\ref{eq:asymptotic_relation}). The one-dimensional marginal
    distributions are given in
    Fig.\,\ref{fig:KIC6928997_distribution}. The $\chi^2_{\mathrm{grid}}$ levels are indicated by the various colourbars. \textit{Top left:} Correlation map
    of the marginal bi-dimensional parameter space ($\Delta \Pi_{1,\rm
      asym}$,\,$q$). The darkest shade indicates the most likely
    solution. \textit{Bottom left:} Correlation map of the the marginal
    bi-dimensional parameter space ($\Delta \Pi_{1,\rm
      asym}$,\,$\epsilon_{\mathrm{g}}$). A strong correlation is observed
    between both parameters. \textit{Bottom right:} Correlation map of the the
    marginal bi-dimensional parameter space ($q$,\,$\epsilon_{\mathrm{g}}$). Correlation maps for the two other stars in our sample are given in the Appendix.}
\label{fig:KIC6928997_grid}%
\end{figure*}

Figure\,\ref{fig:KIC6928997_deltaP} illustrates that the Lorentzian profile
  captures the mixed nature of the dipole modes reasonably well for
  KIC\,6928997. As such, the value of $\Delta \Pi_{1, \rm emp}$ can be used as a
  better starting point, compared to $\overline{\Delta P}$, for the more
  elaborate tri-dimensional parameter search. From the Monte Carlo
  analysis we were able to determine the uncertainties for $\Delta \Pi_{1, \rm
    emp}$, which had a Gaussian-like distribution for both KIC\,6928997 and
  KIC\,10593078. For KIC\,6762022, however, the profile indicated a bi-modal
  distribution, leading to a much larger uncertainty on $\Delta \Pi_{1, \rm
    emp}$. This is caused by a strongly gravity-dominated mixed dipole mode. Performing perturbations of the two corresponding frequencies will often give a $\Delta P$ value larger than the optimal $\Delta \Pi_{1, \rm emp}$, producing the bi-modality of the Monte Carlo distribution.

\begin{figure}[t!]
\centering
\includegraphics[width=\columnwidth, height = 0.33\textheight]{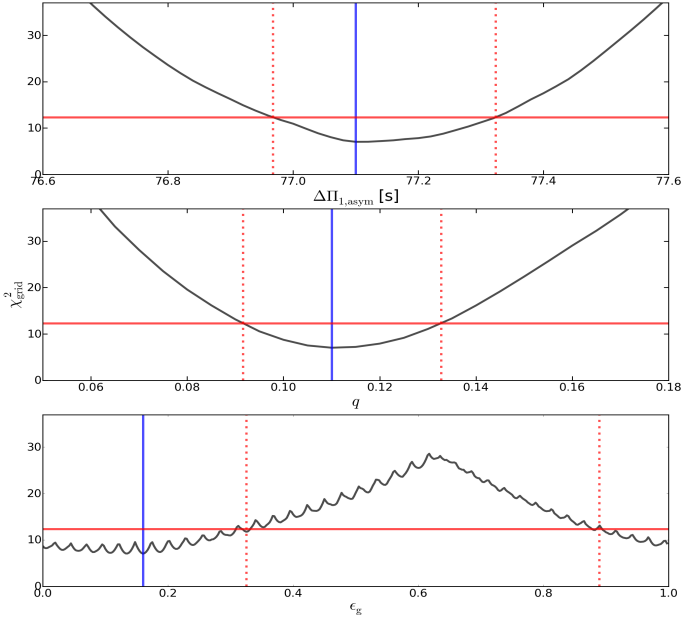}%
\caption{One dimensional marginal distributions for each parameter
    considered in the asymptotic relation for KIC\,6928997, centred on the
    optimal solution, which is marked in blue. The $\chi^2_{\mathrm{95\%}}$ value is marked by the solid
    red line. The upper (lower) boundaries for the uncertainty on the individual
    parameters are given by the red dotted lines. The marginal distribution for each parameter over the full tri-dimensional grid is given in Fig.\,\ref{fig:KIC6928997_distribution} in the Appendix.}
\label{fig:KIC6928997_uncertainty}%
\end{figure}

Figures \ref{fig:KIC6928997_grid} and \ref{fig:KIC6928997_uncertainty} show the different marginal distributions
  for the mixed modes derived from the asymptotic relation for KIC\,692897 (the same figures are given in the Appendix for the two remaining red giants of the sample) . We
  accept the $\Delta \Pi_{1, \rm asym}$ values derived with this method as our
  final values for $\Delta \Pi_{1}$. Studying the different marginal
  distributions enabled use to discuss the behaviour of the asymptotic relation
  in more detail.

\begin{figure}[t!]
\centering
\includegraphics[width=\columnwidth, height = 0.33\textheight]{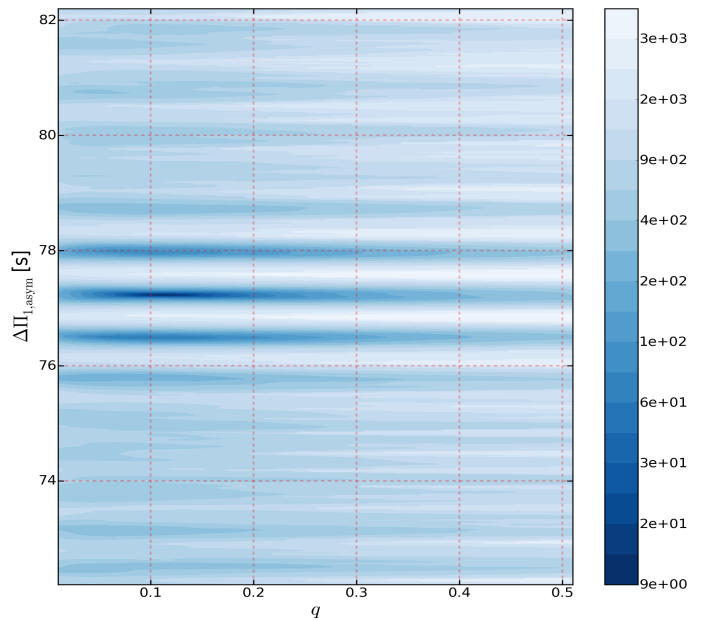}%
\caption{Correlation maps indicating the $\chi^2_{\mathrm{grid}}$ as a colour map for KIC\,6928997, as deduced from the asymptotic relation Eq.\,(\ref{eq:asymptotic_relation}) and assuming $\epsilon_{\mathrm{g}}$ is fixed at zero.}%
\label{fig:KIC6928997_dpq}%
\end{figure}

The correlation maps show that there is a significant correlation
  between $\Delta \Pi_{1, \rm asym}$ and the phase offset 
  $\epsilon_{\mathrm{g}}$ (bottom left panel Fig.\,\ref{fig:KIC6928997_grid}). This was anticipated from looking at
  Eq.\,(\ref{eq:asymptotic_relation}), since both parameters are present within
  the tangent-function. As such, it is possible to have the same mixed-mode frequencies for
  various $\Delta \Pi_{1, \rm asym}$ combined with appropriate
  $\epsilon_{\mathrm{g}}$ values. Also, fixing $\epsilon_{\mathrm{g}}$ to one unique
  value prohibits capturing the complete behaviour of the asymptotic
  relation. This is shown in Fig.\,\ref{fig:KIC6928997_dpq}, where we indicate the correlation map between $\Delta \Pi_{1, \rm asym}$ and $q$, assuming $\epsilon_{\mathrm{g}}=0$. Since the phase offset is assumed to be constant, the frequency space of the mixed modes is not fully sampled. Thus, the colour map mimics a correlation between $\Delta \Pi_{1, \rm asym}$ and $q$ and produces a multi-modal behaviour in the marginal $\Delta \Pi_{1, \rm asym}$ distribution. No other significant correlations are observed between the other
  parameters in the tri-dimensional parameter space nor between the coupling factor, $q$, and the value of the tangent term within the asymptotic relation, Eq.\,(\ref{eq:asymptotic_relation}).
  However, the marginal distribution of $\epsilon_{\mathrm{g}}$ is sensitive to the sampling rate along the $\Delta \Pi_{1, \rm asym}$ axis. The larger the difference between consecutive $\Delta \Pi_{1, \rm asym}$ values, the stronger the wiggles are in the bottom panel of Fig.\,\ref{fig:KIC6928997_uncertainty}. These wiggles are understood as the influence of the correlation between $\Delta \Pi_{1, \rm asym}$ and $\epsilon_{\mathrm{g}}$ on the chosen sampling rates. Indeed, performing a first order perturbation analysis of Eq.\,\ref{eq:asymptotic_relation}, assuming a constant $q$ and $\nu_{\rm m}$, leads to the relation
\begin{equation}
\delta(\epsilon_{\mathrm{g}}) = \left|\frac{\delta(\Delta\Pi_{1, \rm asym})}{\Delta\Pi_{1, \rm asym}^{2} \nu_{\rm m} }\right| \ \mathrm{,}
\label{eq:asymptotic_relation_perturbation}
\end{equation}
where $\delta(\epsilon_{\mathrm{g}})$ and $\delta(\Delta\Pi_{1, \rm asym})$ are the perturbations on $\epsilon_{\mathrm{g}}$ and $\Delta\Pi_{1, \rm asym}$, respectively. Considering $\delta(\Delta\Pi_{1, \rm asym})$ as our chosen sampling rate of $\Delta\Pi_{1, \rm asym}$ indicates that we inherently sample our $\epsilon_{\mathrm{g}}$ axis. Using the accepted $\Delta\Pi_{1, \rm asym}$, approximating $\nu_{\rm m}$ as $\nu_{\rm max}$, and taking the chosen sampling rate $\delta(\Delta\Pi_{1, \rm asym}) = 0.02$\,s for KIC\,6928997 results in a $\delta(\epsilon_{\mathrm{g}}) = 0.028$, which is similar to the size of the wiggles observed in the bottom panel of Fig.\,\ref{fig:KIC6928997_uncertainty}. A smaller sampling rate along $\Delta \Pi_{1, \rm asym}$ would provide a smoother marginal profile at a computational expense. However, it would be unlikely that this smoother distribution  behaves substantially different. 

Second, the inclusion of the phase offset $\epsilon_{\mathrm{g}}$ as a
  variable parameter permitted us to determine correct confidence interval for
  each parameter. Both $\Delta \Pi_{1, \rm asym}$ and $q$ are strongly confined,
  unlike $\epsilon_{\mathrm{g}}$. The confidence interval for
  $\epsilon_{\mathrm{g}}$ spans a large portion of the parameter space. We mark
  the boundaries for the various confidence intervals in
  Fig.\,\ref{fig:KIC6928997_uncertainty}. Nevertheless, the inclusion of the phase offset is mandatory if one wants to study $\Delta \Pi_{1, \rm asym}$ in detail, due the above-mentioned correlation. In the present analysis, $\epsilon_{\mathrm{g}}$ does not in itself provide very useful information, owing to its large confidence interval, but must be  accounted for in the study of the parameters which are of interest, $\Delta \Pi_{1, \rm asym}$ and $q$. Despite this behaviour of $\epsilon_{\mathrm{g}}$, we did not lose any information, since $\epsilon_{\mathrm{g}}$ was artificially set to zero in previous studies, but did gain more understanding of $\Delta \Pi_{1, \rm asym}$.

Lastly, we note that the marginal distribution of $\Delta \Pi_{1, \rm
    asym}$ looks significantly different for both KIC\,6762022 and KIC\,10593078
  compared to KIC\,6928997 (see figures in Appendix). The distributions of the former show
  a different shape around the minimum $\chi^2_{\mathrm{grid}}$ value,
  resembling a local and a global minimum. A possible explanation is the
  presence of buoyancy glitches, giving rise to slightly different $\Delta
  \Pi_{1, \rm asym}$ values for different radial-mode orders. A detailed study
  of the dipole mixed-mode frequencies per radial-mode order is of interest, but
  is out of the scope of the current work.

\begin{figure}[t!]
\centering
\includegraphics[width=\columnwidth, height = 0.33\textheight]{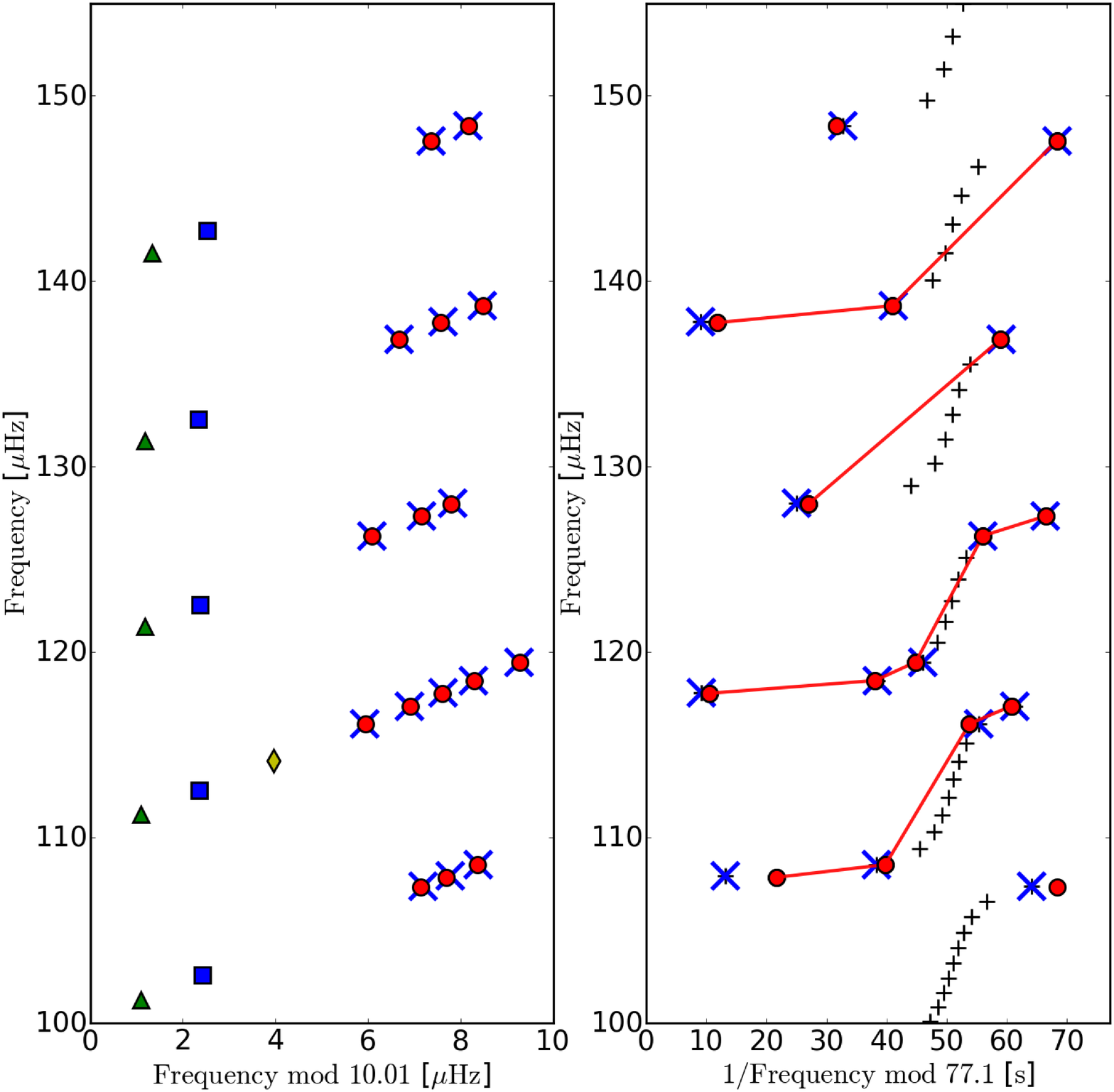}%
\caption{\'Echelle diagrams comparing the solution of the asymptotic relation
  ($\Delta \Pi_{1,\rm asym} = 77.1 \ \mathrm{s}$, $q = 0.11$ and $\epsilon_{\mathrm{g}} = 0.16$) with extracted modes for
  KIC\,6928997. \textit{Left}: a frequency \'echelle diagram, comparing the
  frequencies of the asymptotic relation to those from the mode extraction. The
  extracted radial, dipole, quadrupole and octupole modes are indicated by blue
  squares, red dots, green triangles, and yellow diamonds, respectively. The
  frequencies of the dipole mixed modes obtained from the asymptotic relation,
  and having an observational counterpart, are indicated by blue
  crosses. \textit{Right}: a period \'echelle spectrum for the dipole modes,
  showing the same comparison as in the left panel with the same colour
  coding. The black '+' indicate the frequencies for the mixed modes obtained
  from the asymptotic relation without any observational counterparts.}
			\label{fig:KIC6928997_asymptechelle}
\end{figure}

We construct frequency and period \'echelle diagrams (such as
Fig.\,\ref{fig:KIC6928997_asymptechelle}) for further visual comparison
between the frequencies of the dipole mixed modes extracted from the PSD and
those calculated with the asymptotic relation $\nu_{\rm m}$. Only very small
differences between the extracted dipole mixed modes and the corresponding
$\nu_{\rm m}$ are seen in the frequency \'echelle diagram. In addition, the
period \'echelle spectrum captures the behaviour of the mixed modes very well,
confirming that the extracted modes indeed show a mixed behaviour, making the
asymptotic relation appropriate. 

Differences between values for \DP derived by the two methods are
  slightly larger than the uncertainties and more pronounced for the confidence
  intervals of \DP themselves. Uncertainties determined from the tri-dimensional
  parameter space are significantly smaller. In addition, the
  asymptotic relation provides information related to the strength of the
  coupling between the different propagation zones and a slight hint for the
  phase offset required for asymptotic theory. Our results determined
  by means of the asymptotic relation agree with those in the literature, yet
  are slightly different since we considered $\epsilon_{\mathrm{g}}$ to be a free
  parameter. The major contrast is witnessed for the uncertainties for $\Delta
  \Pi_{1, \rm asym}$. They range between those proposed by
  \citet{2012A+A...540A.143M} and \citet{2014A+A...572L...5M} and are
  asymmetric. 
  
\begin{table*}[t!]
\caption{Results for the period spacing determined from the extracted dipole frequencies of Q0-Q17 \Kepler data.}
\centering
\tabcolsep=12pt
\begin{tabular}{lcccccc}
\hline
\hline
KIC	& $\overline{\Delta P}$&Evolutionary &$\Delta \Pi_{1, \rm emp}$&$\Delta \Pi_{1,\rm asym}$&$q$&$\epsilon_{\mathrm{g}}$\\
							&&state&Empirical&\multicolumn{3}{c}{Asymptotic}\\
							&[s]&[s]&[s]&[s]&[ ]&[ ]\\
\hline
6928997			&$53.1 \pm 11.2$&RGB&$	78.3		_{-3.3}^{+4.2}		$&$	77.1	0	_{-0.13}^{+0.22}	$&$	0.111	^{+0.023}	_{-0.018}	$&$	0.160	_{-0.271}^{+0.165}$\rule{0pt}{3ex}\\
6762022			&$210.3 \pm 19.1$&RC&$	261.9	_{-18.4}^{+25.9~a}	$&$	259.08	_{-1.63}^{+1.19}	$&$	0.240	^{+0.091}	_{-0.063}	$&$	0.835	_{-0.420}^{+0.580}$\rule{0pt}{3ex}\\
10593078			&$53.7 \pm 14.3$&RGB&$	81.8		_{-1.0}^{+1.2}		$&$	82.48	_{-0.83}^{+0.47}	$&$	0.130	^{+0.073}	_{-0.054}	$&$	0.755	_{-0.326}^{+0.608}$\rule{0pt}{3ex}\\
\hline
\end{tabular}
\tablefoot{Confidence intervals are 95\,$\%$ intervals, except the 68\,$\%$ for $\overline{\Delta P}$. $a$: the Monte Carlo routine resulted in a strong bi-model distribution. The uncertainty indicated here is when the correct peak is considered.}
\label{tab:results_DP}
\end{table*}

\section{Conclusions}
\label{sect:Discussion}

In this work, we analysed the dipole mixed mode period spacing of
three red giants observed by the NASA \kepler space telescope. Two of them are
in the evolutionary stage of hydrogen shell burning, while the third one was
confirmed to be in the more advanced helium core-burning phase. 
We determine the value for \DP, the \textit{g}-mode asymptotic period spacing,
according to two different approaches. First, we used an empirical fit
to the observed period spacings. Next, the description of the asymptotic
relation is used to study the tri-dimensional parameter space of $\Delta \Pi_{1,
  \rm asym}$, $q$, and $\epsilon_{\mathrm{g}}$. We were able to 
determine realistic confidence intervals for both the asymptotic period spacing and the dimensionless coupling factor. The
  phase offset $\epsilon_{\mathrm{g}}$, however, remains ill defined due its large confidence interval. Yet, it is only by considering $\epsilon_{\mathrm{g}}$ as a variable parameter and using marginal distributions that the determination of a confidence interval for the asymptotic period spacing is simplified, because a fixed $\epsilon_{\mathrm{g}}$ provides a multi-modal behaviour in the $\chi^2$ landscape.

The two approaches have very different computational efficiencies, but lead to
compatible results. Our conclusion is that, when analysing large samples of
stars, particular attention has to be given to the techniques adopted to
estimate the value of \DP and meaningful uncertainties are needed in order to be
able to perform a reliable comparison between observations and stellar
models. The results obtained in this work, allowing for a varying $\epsilon_{\rm g}$, provide reliable uncertainty estimates, which are in between those quoted by \citet{2012A+A...540A.143M} and \citet{2014A+A...572L...5M}, i.e. meaningful estimates of the relative uncertainty of the period spacing range up to 1\,\%.

Determining the period spacing per radial-mode order constitutes a next step
  onwards, since the marginal distribution of $\Delta \Pi_{1,\rm asym}$
  indicated a substructure around the optimal solution. This would provide
  information about possible structural glitches in the core, which were ignored
  in this work. However, this is only possible if enough dipole mixed modes are
  identified per radial-mode order. Another possibility is to include the large
  frequency separation as a fourth parameter in the study of the asymptotic
  relation. At present, we fixed this value since it was deduced with a very
  high accuracy during the detailed frequency analysis.

\begin{acknowledgements}

We acknowledge the work of the team behind \textit{Kepler}. Funding for the \textit{Kepler} Mission is provided by NASA's Science Mission Directorate. 
B.B. thanks Dr. Rasmus Handberg for helpful discussions.
The research leading to these results has received funding from the Fund for Scientific Research of Flanders (FWO, project G.0728.11).
Funding for the Stellar Astrophysics Centre is provided by The Danish National Research Foundation (Grant DNRF106). The research is supported by the ASTERISK project (ASTERoseismic Investigations with SONG and Kepler) funded by the European Research Council (Grant agreement no.: 267864).
B.B. acknowledges funding by the KU Leuven and the European Union for an Erasmus stay at Aarhus Universiteit.
P.G.B. and R.G. acknowledge the ANR (Agence Nationale de la Recherche, France) program IDEE (n$\degs$ ANR-12-BS05-0008) "Interaction Des Etoiles et des Exoplanetes".
E.C. is funded by the European Community's Seventh Framework Programme (FP7/2007-2013) under grant agreement n$^\circ$312844 (SPACEINN).
The research leading to these results has received funding from the Fund for Scientific Research of Flanders (G.0728.11).
V.S.A. acknowledges support from VILLUM FONDEN (research grant 10118).

\end{acknowledgements}
\bibliographystyle{aa}
\bibliography{Thesis_ADS}

\newpage
\begin{appendix}
\section{Marginal distributions}
\FloatBarrier
Here, we present the various marginal distributions for the two remaining red giants, KIC\,6762022 and KIC\,10593078. They are similar to the figures presented in Sect.\,\ref{sect:DP_results}. In addition, we provide the 1D marginal distributions over the full tri-dimensional grid for each parameter and for each star.

\begin{figure}[!h]
\centering
\includegraphics[width=\columnwidth, height = 0.33\textheight]{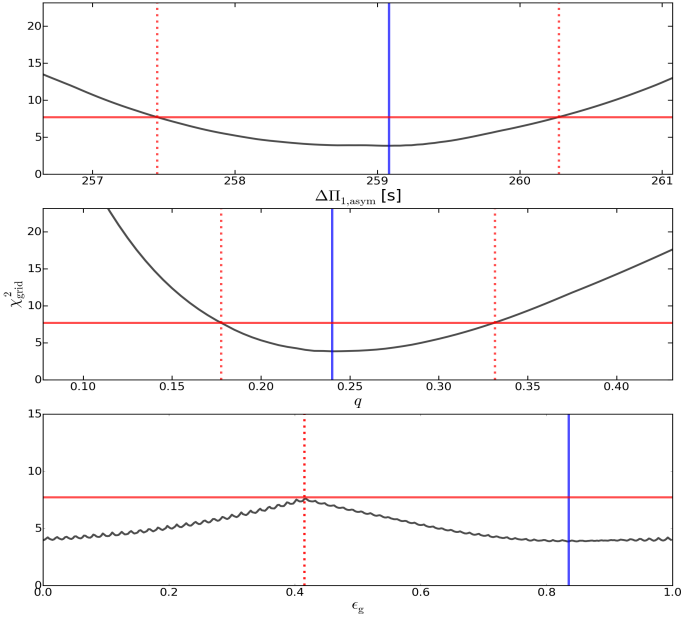}%
\caption{One dimensional marginal distributions for each parameter
    considered in the asymptotic relation for KIC\,6762022, centred at the
    optimal solution. Similar figure as
    Fig.\,\ref{fig:KIC6928997_uncertainty}.}
\label{fig:KIC6762022_uncertainty}%
\end{figure}

\begin{figure}[!h]
\centering
\includegraphics[width=\columnwidth, height = 0.33\textheight]{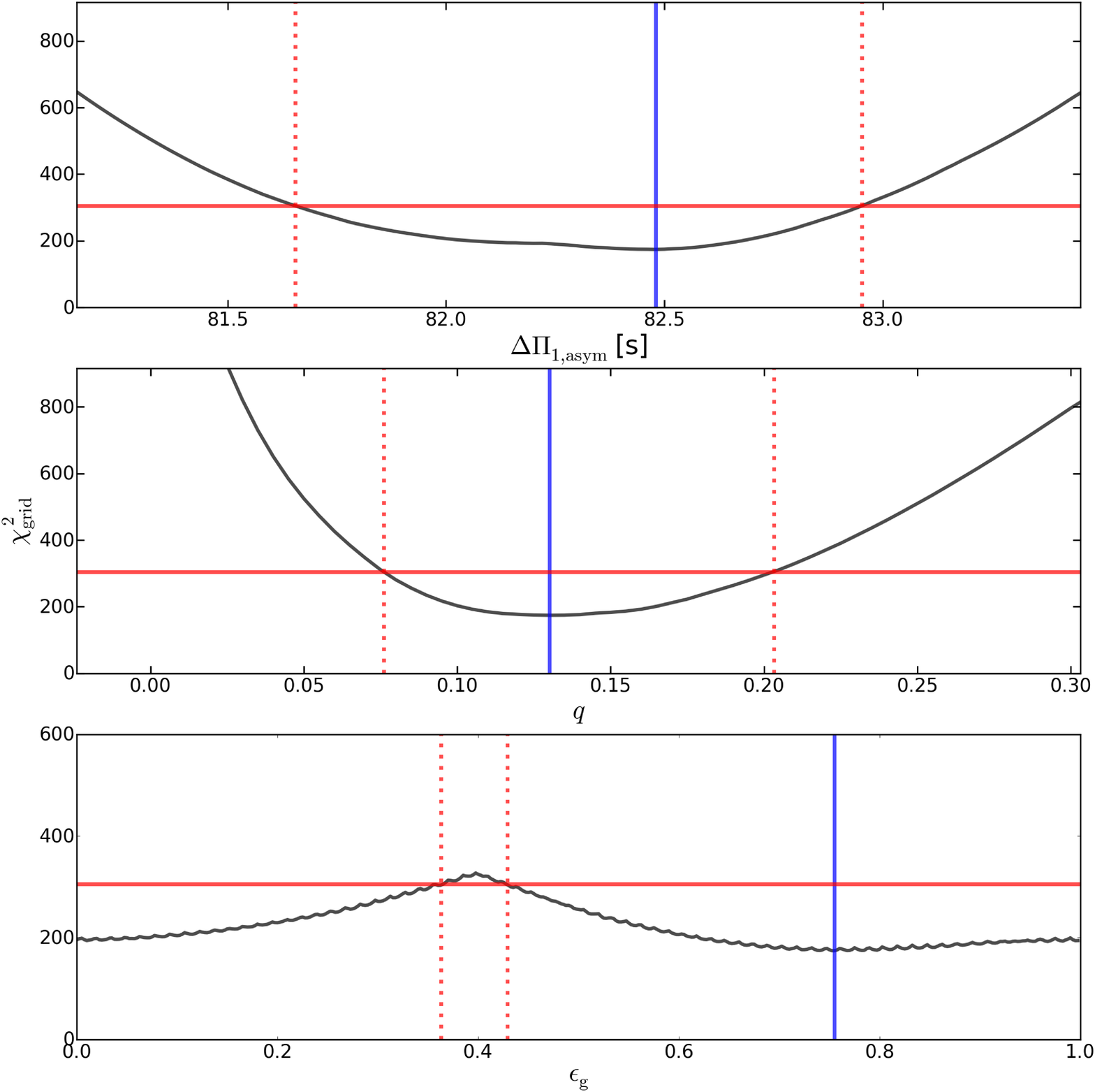}%
\caption{One dimensional marginal distributions for each parameter considered in the asymptotic relation for KIC\,10593078, centred at the optimal solution. Similar figure as Fig.\,\ref{fig:KIC6928997_uncertainty}.}
\label{fig:KIC10593078_uncertainty}%
\end{figure}

\FloatBarrier
\begin{figure*}[!h]
\centering
\includegraphics[width=\textwidth]{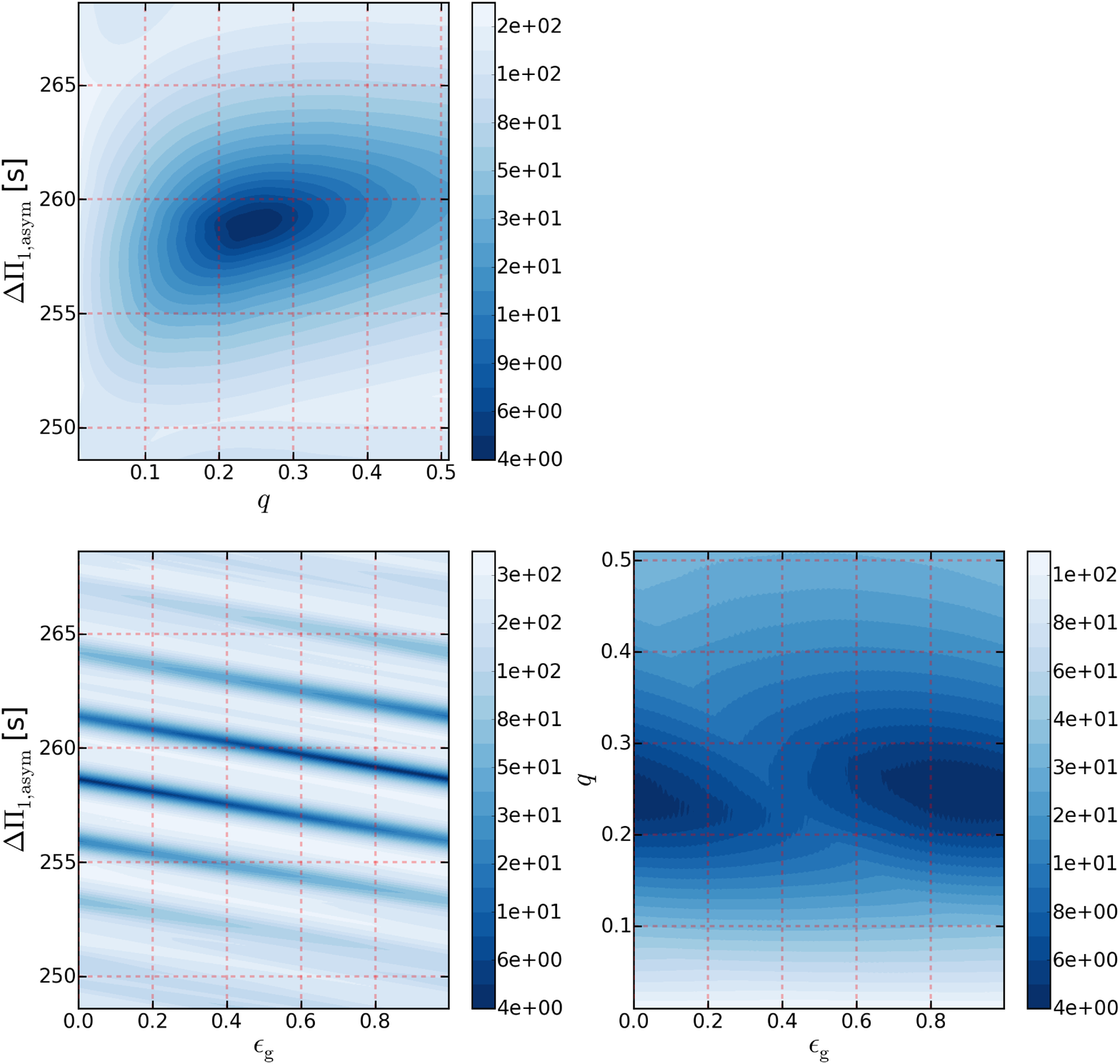}%
\caption{Correlation maps indicating the $\chi^2_{\mathrm{grid}}$ as a colour map for KIC\,6762022, as deduced with the asymptotic relation Eq.\,(\ref{eq:asymptotic_relation}). Similar figure as Fig.\,\ref{fig:KIC6928997_grid}.}
\label{fig:KIC6762022_grid}%
\end{figure*}

\begin{figure*}[!h]
\centering
\includegraphics[width=\textwidth]{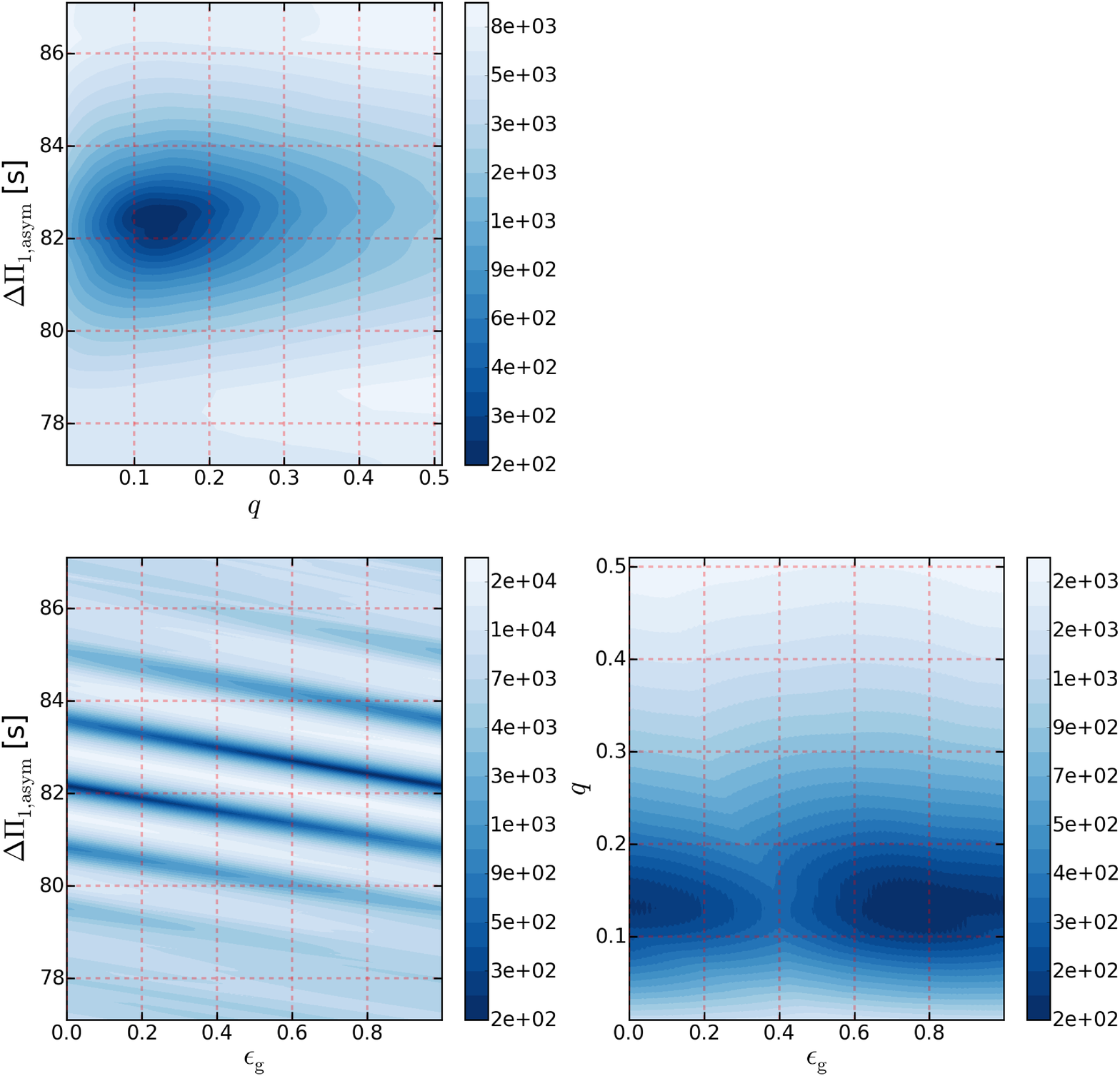}%
\caption{Correlation maps indicating the $\chi^2_{\mathrm{grid}}$ as a
    colour map for KIC\,10593078, as deduced with the asymptotic relation
    Eq.\,(\ref{eq:asymptotic_relation}). Similar figure as
    Fig.\,\ref{fig:KIC6928997_grid}.}
\label{fig:KIC10593078_grid}%
\end{figure*}

\FloatBarrier
\begin{figure}[!h]
\centering
\includegraphics[width=\columnwidth, height = 0.33\textheight]{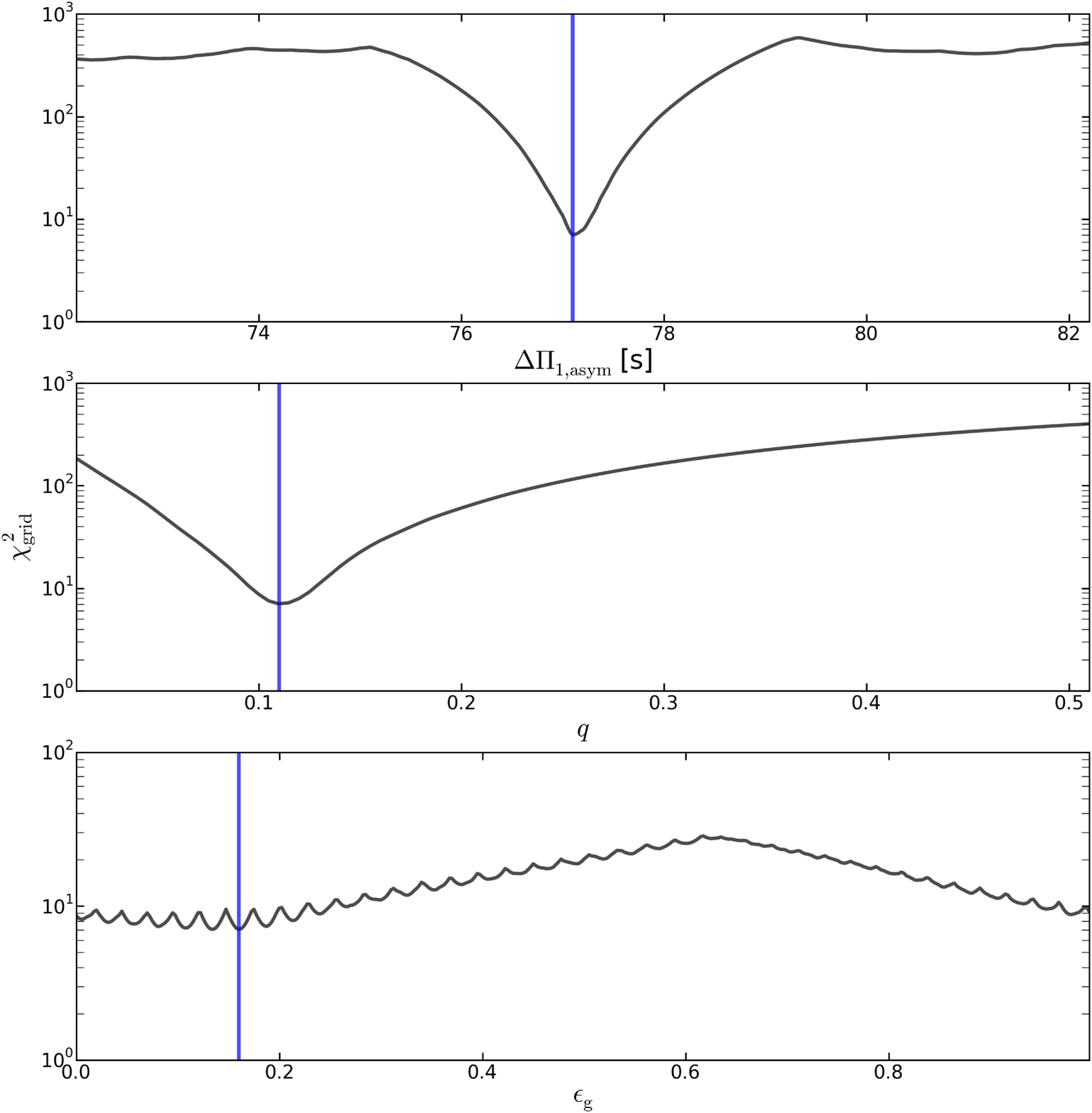}%
\caption{One dimensional marginal distributions for each parameter
    considered in the asymptotic relation for KIC\,6928997. The final values for
    each parameter, corresponding to the best description for the dipole mixed
    modes with the asymptotic relation are marked in blue.}%
\label{fig:KIC6928997_distribution}%
\end{figure}

\begin{figure}[!h]
\centering
\includegraphics[width=\columnwidth, height = 0.33\textheight]{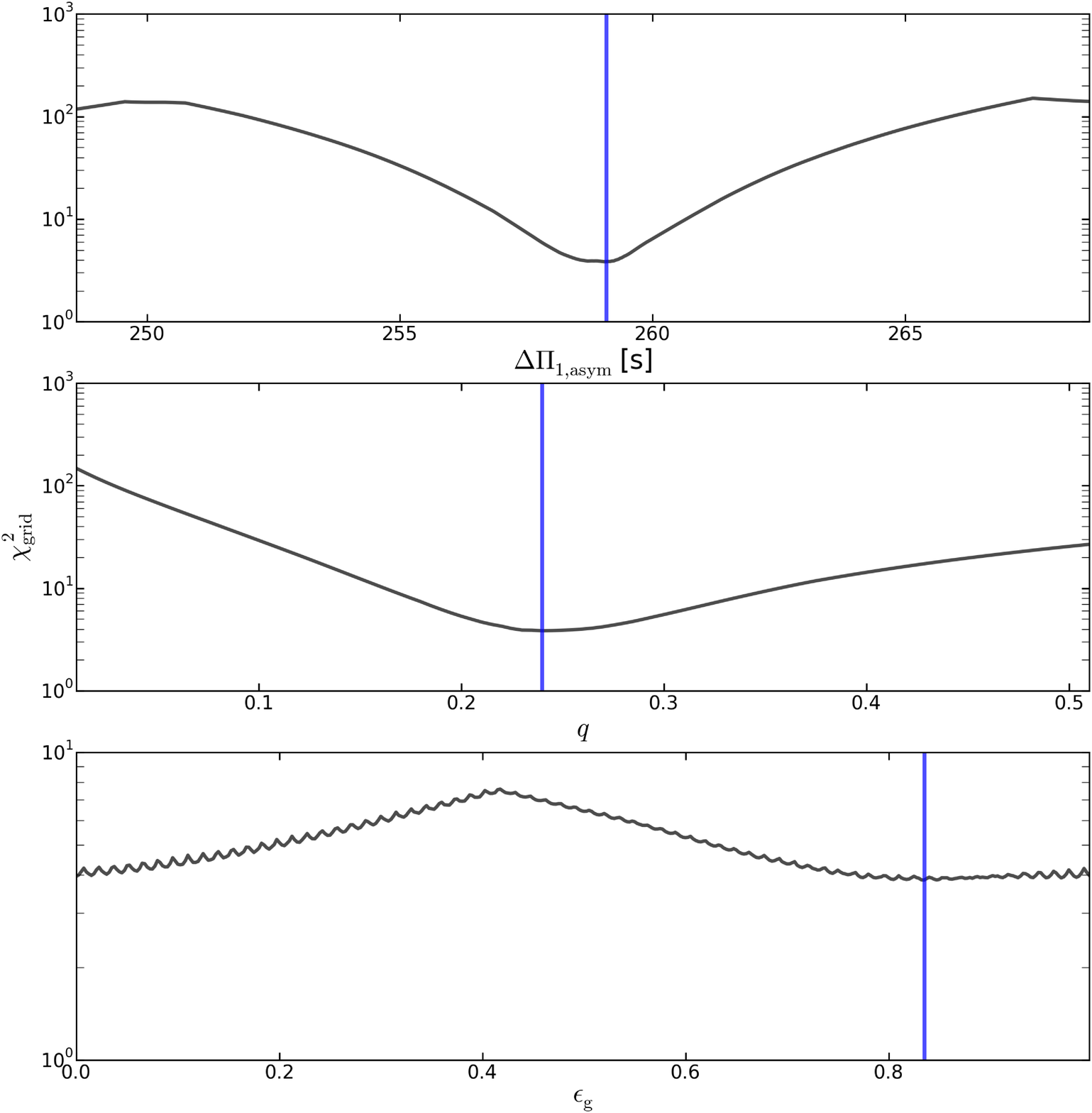}%
\caption{One dimensional marginal distributions for each parameter considered in the asymptotic relation for KIC\,6762022. Similar figure as Fig.\,\ref{fig:KIC6928997_distribution}.}%
\label{fig:KIC6762022_distribution}%
\end{figure}

\begin{figure}[!h]
\centering
\includegraphics[width=\columnwidth, height = 0.33\textheight]{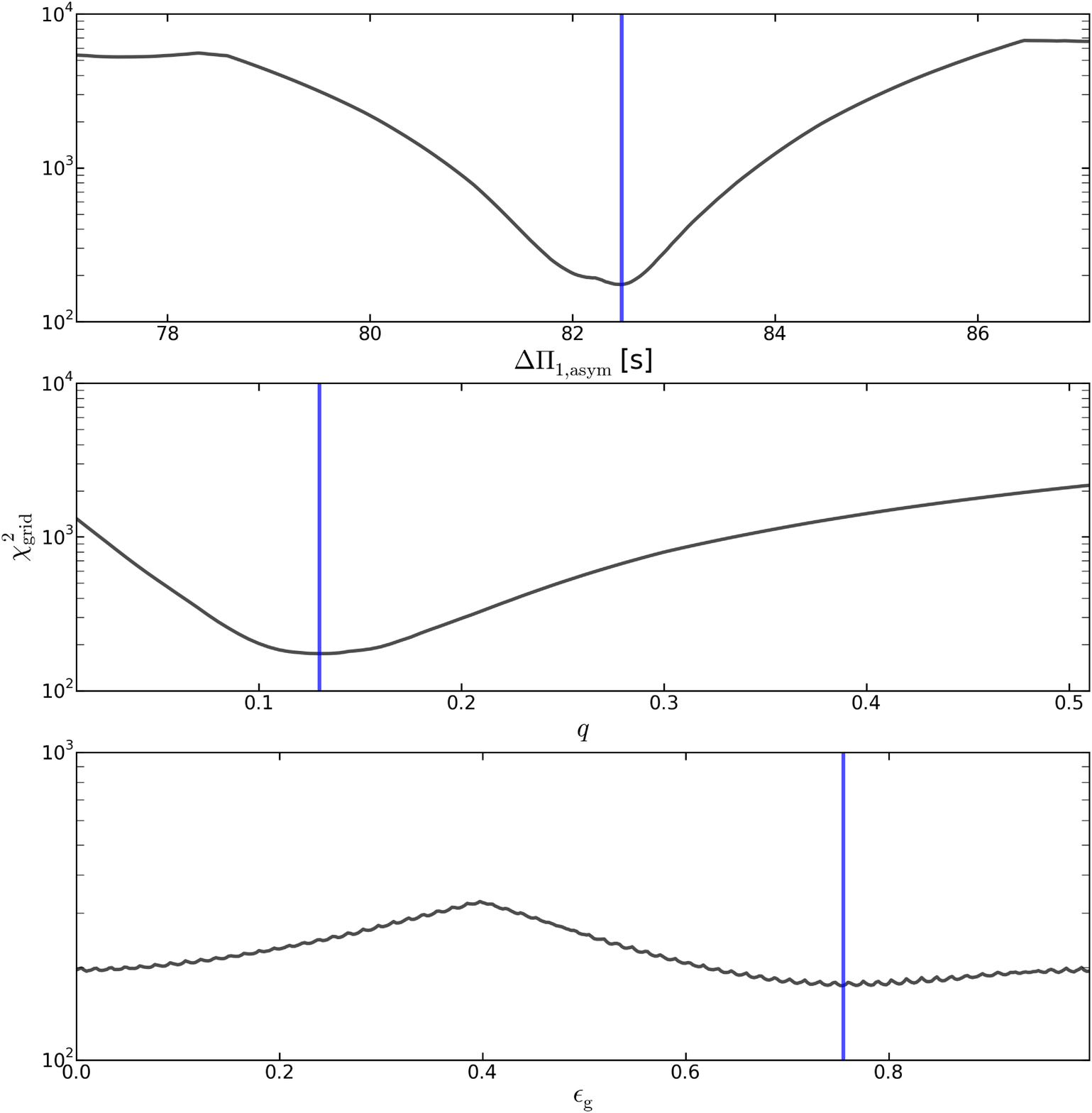}%
\caption{One dimensional marginal distributions for each parameter
    considered in the asymptotic relation for KIC\,10593078. Similar figure as
    Fig.\,\ref{fig:KIC6928997_distribution}.}%
\label{fig:KIC10593078_distribution}%
\end{figure}

\end{appendix}
\end{document}